\documentclass[10pt,english]{article}
\usepackage{graphicx}
\usepackage[T1]{fontenc}
\usepackage[latin9]{inputenc}
\usepackage[letterpaper]{geometry}
\usepackage{float}
\usepackage{amsbsy}
\usepackage{setspace}
\usepackage{amssymb}
\usepackage{esint}
\makeatletter

\floatstyle{ruled}
\newfloat{algorithm}{tbp}{loa}
\floatname{algorithm}{Algorithm}

\usepackage{babel}

\begin{document}

\title{Exact Fourier Spectrum Recovery}

\author{M. Andrecut}

\maketitle
{

\centering ISIS, University of Calgary, Alberta, T2N 1N4, Canada

\centering (email: mandrecu@ucalgary.ca)

} 
\begin{abstract}
Discrete Fourier Transform (DFT) is widely used in signal processing
to analyze the frequencies in a discrete signal. However, DFT fails
to recover the exact Fourier spectrum, when the signal contains frequencies
that do not correspond to the sampling grid. Here, we present an exact
Fourier spectrum recovery method and we provide an implementation
algorithm. Also, we show numerically that the proposed method is
robust to noise perturbations. 
\end{abstract}

\section{Introduction}

Discrete Fourier Transform (DFT) is one of the most important tools
in digital signal processing. DFT is used in both analysis and synthesis
of discrete signals and systems, and has frequent applications in
optics, spectroscopy, magnetic resonance, quantum computing, seismology
and astrophysics \cite{key-1,key-2}. In particular, DFT allows discrete-time
signals and systems to be analyzed in the frequency domain. 

DFT is an invertible linear operator $\mathcal{F}:\mathbb{C^{N}}\rightarrow\mathbb{C}^{N}$,
that transforms a set of complex numbers, defining the signal $z_{n}=z(t_{n})\in\mathbb{C}$
, sampled on the time grid $t_{n}=n\triangle t$, $n=0,...,N-1$,
into another complex set of numbers:\begin{equation}
Z_{m}=\mathcal{F}_{m}[\{z_{n}\}_{n=0}^{N-1}]=N^{-1}\sum_{n=0}^{N-1}z_{n}e^{-2\pi it_{n}\bar{\nu}_{m}},\end{equation}
where $\bar{\nu}_{m}=m/(N\Delta t)$, $m=0,...,N-1$ is the frequency
sampling grid. The inverse DFT is defined as following:\begin{equation}
z_{n}=\mathcal{F}_{n}^{-1}[\{Z_{m}\}_{m=0}^{N-1}]=\sum_{m=0}^{N-1}Z_{m}e^{2\pi it_{n}\bar{\nu}_{m}}.\end{equation}
A major short coming of DFT is that it correctly recovers the Fourier
spectrum only if the signal contains frequencies from the sampling
grid $\{\bar{\nu}_{m}=m/(N\Delta t)\}_{m=0}^{N-1}$ \cite{key-1,key-2}.
In Figure 1 we give a DFT failing recovery example where the signal
has a Fourier spectrum containing $K=8$ components with frequencies
$\nu_{k}$ $(k=0,...,K-1)$, that are not included in the sampling
grid, i.e. $\nu_{k}\notin\{\bar{\nu}_{m}=m/(N\Delta t)\}_{m=0}^{N-1}$.
One can see that the spectrum obtained using DFT does not even resemble
the original frequency spectrum embedded in the signal. The standard
approach to overcome this problem is to sample the signal at a larger
number of points $N$, such that the frequencies in the sampling grid
begin to approximate well the frequencies embedded in the signal.
This approach is obviously not feasible if one can sample the signal
only at a fixed number of points, constrained by the experimental
setting for example. The main question we would like to address here
is how to recover exactly the Fourier spectrum if the signal in question
contains frequencies that do not fall on the sampling grid? Here,
we derive an iterative method which recovers exactly the Fourier spectrum
embedded in the signal, and we provide an implementation algorithm
which is robust to noise perturbations.

\section{Exact Recovery Method}

The goal of the Fourier spectrum recovery problem is to exactly find
the Fourier components $\{Z_{k},\nu_{k}\}_{k=0}^{K-1}$, $K\leq N/2$,
from the signal:\begin{equation}
z_{n}=\sum_{k=0}^{K-1}Z_{k}e^{2\pi it_{n}\nu_{k}},\quad n=0,...,N-1,\end{equation}
including the case when the frequencies $\nu_{k}$ do not fall on
the sampling grid.

We consider an iterative method, and we assume that initially $Z_{k}=0$,
$\nu_{k}=0$, $k=0,...,K-1$, and the residual is $\hat{r}=\left[z_{0},...,z_{N-1}\right]^{T}\in\mathbb{\mathbb{C}}^{N}$.
At every step the method uses a nonlinear optimization strategy to
determine the amplitude $Z_{k}$ and the frequency $\nu_{k}$ corresponding
to the Fourier component $\hat{\psi}(\nu_{k})=\left[e^{2\pi it_{0}\nu_{k}},...,e^{2\pi it_{N-1}\nu_{k}}\right]^{T}\in\mathbb{\mathbb{C}}^{N}$,
which minimize the residual in the least squares sense. A cyclic strategy
is also employed to reassess and correct the $(Z_{k},\nu_{k})$ parameters,
while holding the others fixed. In order to reassess the component
$k$ with the parameters $(Z_{k},\nu_{k})$ we remove it from the
solution, by adding its contribution to the residual:\begin{equation}
\hat{r}\leftarrow\hat{r}+Z_{k}\hat{\psi}(\nu_{k}).\end{equation}
Here, we have introduced the assign operator $variable\leftarrow expression/program$
(re-sets the \textit{$variable$} with the returned value of the\textit{
$expression/program$}). Now, we compute another set of parameters
$(Z_{k},\nu_{k})$ which minimize the residual $\hat{r}$:\begin{equation}
(Z_{k},\nu_{k})\leftarrow\arg\min_{(Z,\nu)}S_{k}(Z,\nu),\end{equation}
where

\begin{equation}
S_{k}(Z,\nu)=\left\Vert \hat{r}-Z\hat{\psi}(\nu)\right\Vert ^{2}=\sum_{n=0}^{N-1}\left[r_{n}-Ze^{2\pi it_{n}\nu}\right]^{2}.\end{equation}
By solving the above minimization problem we find a new set of parameters
$Z_{k}$ and $\nu_{k}$, which will produce the residual:\begin{equation}
\hat{r}\leftarrow\hat{r}-Z_{k}\hat{\psi}(\nu_{k}).\end{equation}
The new residual $\hat{r}$ will be smaller or at least equal with
the previous one, due to the intrinsic convergence mechanism of the
method (as shown below). The worst case would be when the values of
the reassessed parameters $(Z_{k},\nu_{k})$ are identical with the
initial ones and no correction can be made. Obviously, this procedure
can be repeated cyclically for each component until the norm of the
residual becomes zero, or smaller than some prescribed positive threshold:
$\left\Vert \hat{r}\right\Vert <\varepsilon_{r}\geqq0$. When the
signal is contaminated with Gaussian noise with zero mean and standard
deviation $\sigma$, the stopping threshold should be comparable with
the standard deviation of the noise, i.e. $\varepsilon_{r}\simeq\sigma$.

The major difficulty of the method is solving the successive nonlinear
minimization problems (5). The functions $S_{k}(Z,\nu)$ are highly
nonlinear, with complex arguments and multiple local minima, which
makes them very difficult for the existing algorithms \cite{key-3}.
Here, we develop a new iterative algorithm to solve the nonlinear
minimization problem. First, we need to find an initial estimate $\nu_{k}$.
This is is done simply by solving the problem:\begin{equation}
\nu_{k}\leftarrow\arg\max_{\bar{\nu}_{m}}\left|\left\langle \hat{r},\hat{\psi}(\bar{\nu}_{m})\right\rangle \right|,\end{equation}
where $\left\langle .,.\right\rangle $ is the standard inner product
operator in the complex Hilbert space, and $\bar{\nu}_{m}$ are the
frequencies from the sampling grid $\{\bar{\nu}_{m}=m/(N\Delta t)\}_{m=0}^{N-1}$.
We also notice that the amplitude $Z_{k}$ and the frequency $\nu_{k}$
can be {}``decoupled'', in the sense that for a given $\nu_{k}$,
one can easily calculate an estimate of $Z_{k}$ as the projection
of $\hat{r}$ on $\hat{\psi}(\nu_{k})$:\begin{equation}
Z_{k}=N^{-1}\left\langle \hat{r},\hat{\psi}(\nu_{k})\right\rangle .\end{equation}
This estimate of $Z_{k}$ guarantees the decrease of the residual
(7), since we have:\[
\left\Vert \hat{r}-Z_{k}\hat{\psi}(\nu_{k})\right\Vert ^{2}=\left\langle \hat{r}-Z_{k}\hat{\psi}(\nu_{k}),\hat{r}-Z_{k}\hat{\psi}(\nu_{k})\right\rangle =\]
\[
\left\langle \hat{r},\hat{r}\right\rangle -Z_{k}^{*}\left\langle \hat{r},\hat{\psi}(\nu_{k})\right\rangle -Z_{k}\left\langle \hat{\psi}(\nu_{k}),\hat{r}\right\rangle +\left|Z_{k}\right|^{2}\left\langle \hat{\psi}(\nu_{k}),\hat{\psi}(\nu_{k})\right\rangle =\]
\begin{equation}
\left\Vert \hat{r}\right\Vert ^{2}-N\left|Z_{k}\right|^{2}-N\left|Z_{k}\right|^{2}+N\left|Z_{k}\right|^{2}\leq\left\Vert \hat{r}\right\Vert ^{2}.\end{equation}
Reciprocally, $\hat{\psi}(\nu_{k})$ is orthogonal to the residual
(7):\[
\left\langle \hat{\psi}(\nu_{k}),\hat{r}-Z_{k}\hat{\psi}(\nu_{k})\right\rangle =\]
\[
\left\langle \hat{\psi}(\nu_{k}),\hat{r}\right\rangle -Z_{k}^{*}\left\langle \hat{\psi}(\nu_{k}),\hat{\psi}(\nu_{k})\right\rangle =\]
\begin{equation}
Nc_{k}^{*}(t)-Nc_{k}^{*}(t)=0.\end{equation}

Now, let us assume that $\triangle\nu_{k}$ is the unknown correction
to $\nu_{k}$ in the next iteration step. Therefore, we have:\begin{equation}
\hat{\psi}(\nu_{k}+\triangle\nu_{k})\simeq\hat{\psi}(\nu_{k})+\frac{d\hat{\psi(\nu_{k})}}{d\nu}\triangle\nu_{k},\end{equation}
with\begin{equation}
\frac{d\hat{\psi(\nu_{k})}}{d\nu}=2\pi i\left[t_{0}e^{2\pi i\nu_{k}t_{0}},...,t_{N-1}e^{2\pi i\nu_{k}t_{N-1}}\right]^{T},\end{equation}
and from here we easily find: \begin{equation}
\triangle\nu_{k}=\left\Vert Z_{k}\frac{d\hat{\psi}(\nu_{k})}{d\nu}\right\Vert ^{-2}\left\langle \hat{r}-Z_{k}\hat{\psi}(\nu_{k}),Z_{k}\frac{d\hat{\psi}(\nu_{k})}{d\nu}\right\rangle .\end{equation}
Again, this estimate of the correction $\triangle\nu_{k}$ guarantees
a further decrease of the residual:\begin{equation}
\hat{r}\leftarrow\hat{r}-Z_{k}\hat{\psi}(\nu_{k})-Z_{k}\frac{d\hat{\psi}(\nu_{k})}{d\nu}\triangle\nu_{k},\end{equation}
 since we have:\[
\left\Vert \hat{r}-Z_{k}\hat{\psi}(\nu_{k})-Z_{k}\frac{d\hat{\psi}(\nu_{k})}{d\nu}\triangle\nu_{k}\right\Vert ^{2}=\]
\begin{equation}
\left\Vert \hat{r}-Z_{k}\hat{\psi}(\nu_{k})\right\Vert ^{2}-\triangle\nu_{k}^{2}\left\Vert Z_{k}\frac{d\hat{\psi}(\nu_{k})}{d\nu}\right\Vert ^{2}\leq\left\Vert \hat{r}-Z_{k}\hat{\psi}(\nu_{k})\right\Vert ^{2}.\end{equation}
Reciprocally, the direction $Z_{k}\frac{d\hat{\psi(\nu_{k})}}{d\nu}$
is also orthogonal to the next residual:\[
\left\langle Z_{k}\frac{d\hat{\psi}(\nu_{k})}{d\nu},\hat{r}-Z_{k}\hat{\psi}(\nu_{k})-Z_{k}\frac{d\hat{\psi}(\nu_{k})}{d\nu}\triangle\nu_{k}\right\rangle =\]
\[
\left\langle Z_{k}\frac{d\hat{\psi}(\nu_{k})}{d\nu},\hat{r}-Z_{k}\hat{\psi}(\nu_{k})\right\rangle -\triangle\nu_{k}\left\langle Z_{k}\frac{d\hat{\psi}(\nu_{k})}{d\nu},Z_{k}\frac{d\hat{\psi}(\nu_{k})}{d\nu}\right\rangle =\]
\begin{equation}
\triangle\nu_{k}\left\Vert Z_{k}\frac{d\hat{\psi}(\nu_{k})}{d\nu}\right\Vert ^{2}-\triangle\nu_{k}\left\Vert Z_{k}\frac{d\hat{\psi}(\nu_{k})}{d\nu}\right\Vert ^{2}=0.\end{equation}
After $\triangle\nu_{k}$ is determined, we update the frequency of
the current component using $\nu_{k}\leftarrow\nu_{k}+Re\{\triangle\nu_{k}\}$
(since the frequency must be a real number) and the residual: $\hat{r}\leftarrow\hat{r}-Z_{k}\hat{\psi}(\nu_{k})$,
and we perform a new iteration. The iterations stop when $\left|\triangle\nu_{k}\right|\leq\varepsilon_{\nu}$,
where $0\leq\varepsilon_{\nu}\ll1$ is a small acceptable error, or
when the number of iterations exceed a maximum number. Therefore,
at every step, the method finds two directions $\hat{\psi}(\nu_{k})$
and respectively $Z_{k}\frac{d\hat{\psi}(\nu_{k})}{d\nu}$, which
are orthogonal to the next residual. The parameters $(Z_{k},\nu_{k})$
are the projection of the current residual on these two directions.
Thus, as a consequence of the above proof, the method is guaranteed
to converge. The pseudo-code of the Exact Fourier Spectrum Recovery
(EFSR) algorithm is given in Algorithm 1. 

\begin{algorithm}
\caption{Exact Fourier Spectrum Recovery (EFSR) method}

$\hat{z}$; signal

$K$; number of components to be recovered

$\varepsilon_{r}$; admissible residual level

$\varepsilon_{\nu}$; admissible frequency error

$S$; maximum number of correction cycles 

$T$; maximum number of iterations per cycle

$\hat{r}\leftarrow\hat{z}$; initial residual

$s\leftarrow0$; cycle index

do\{

\quad{}$k\leftarrow\mathrm{mod}(s,K)$; 

\quad{}$\hat{r}\leftarrow\hat{r}+Z_{k}\hat{\psi}(\nu_{k})$;

\quad{}$\nu_{k}\leftarrow\arg\max_{\bar{\nu}_{m}}\left|\left\langle \hat{r},\hat{\psi}(\bar{\nu}_{m})\right\rangle \right|$;

\quad{}$t\leftarrow0$;

\quad{}do\{

\quad{}\quad{}$Z_{k}\leftarrow N^{-1}\left\langle \hat{r},\hat{\psi}(\nu_{k})\right\rangle $;

\quad{}\quad{}$\triangle\nu_{k}\leftarrow\left\Vert Z_{k}\frac{d\hat{\psi}(\nu_{k})}{d\nu}\right\Vert ^{-2}\left\langle \hat{r}-Z_{k}\hat{\psi}(\nu_{k}),Z_{k}\frac{d\hat{\psi}(\nu_{k})}{d\nu}\right\rangle $;

\quad{}\quad{}$\nu_{k}\leftarrow\nu_{k}+Re\{\triangle\nu_{k}\}$;

\quad{}\quad{}$t\leftarrow t+1$;

\quad{}\quad{}\}while($\left|\triangle\nu_{k}\right|>\varepsilon_{\phi}$
and $t<T$)

\quad{}$\hat{r}\leftarrow\hat{r}-Z_{k}\hat{\psi}(\nu_{k})$;

\quad{}$s\leftarrow s+1$;

\quad{}\}while($\left\Vert \hat{r}\right\Vert >\varepsilon_{r}$
and $\left\lfloor s/K\right\rfloor <S$)

return $\{Z_{k},\nu_{k}\}_{k=0}^{K-1}$
\end{algorithm}

In the example form Figure 1 one can see that DFT cannot recover the
spectrum at all, while the EFSR method recovers exactly all the components
in the Fourier spectrum ($\varepsilon_{\nu}=\varepsilon_{r}=10^{-6}$,
$T=10^{2}$, $S=K^{2}$) in about $s=10$ iteration cycles. We should mention that the EFSR method will
recover exactly all the cases when the frequencies from the spectrum
are well separated among them by at least the Nyquist limit, which
is the highest frequency that can be coded at a given sampling rate
$\triangle t$ in order to be able to fully reconstruct the signal,
i.e. $\triangle\nu=2.0/(N\triangle t)$. Thus, for a successful recovery
we must have: $\left|\nu_{k}-\nu_{j}\right|\geq\triangle\nu$, for
$k\neq j$, $k,j=0,1,...,K-1$, $K\leq N/2$. 

In order to investigate the effect of the noise perturbation we consider
the perturbed signal $z_{n}+\eta_{n}$, where $\eta_{n}$ is the measurement
noise, i.e. a random variable normal distributed with mean $\left\langle \eta_{n}\right\rangle =0$
and standard deviation $\sigma$. Also, as a measure of noise contamination
we use the signal to noise ratio defined as:\begin{equation}
SNR=\left(\frac{RMS_{signal}}{RMS_{noise}}\right)^{2}\simeq\sigma^{-2}RMS_{signal}^{2},\end{equation}
where $RMS$ is the root mean square.

In Figure 2 we give an example with the signal to noise ratio: $SNR=10$.
One can see that the EFSR method still recovers the spectrum almost
exactly. In this case the iterations were stopped when the norm of
the residual becomes comparable with the standard deviation $\sigma$
of the noise in the measured data, i.e. when: $\left\Vert \hat{r}\right\Vert \leq\sigma$,
and thus $\varepsilon_{r}=\sigma$. In order to estimate the recovery
capabilities of the EFSR method in the presence of noise, we define
the following average relative errors between the recovered spectrum
$\{Z_{k}^{(r)},\nu_{k}^{(r)}\}_{k=0}^{K-1}$ and the original spectrum
$\{Z_{k},\nu_{k}\}_{k=0}^{K-1}$:\begin{equation}
\varepsilon_{Z}=\frac{1}{K}\sum_{k=0}^{K-1}\frac{\left|Z_{k}^{(r)}-Z_{k}\right|}{\left|Z_{k}\right|}\times100\%.\end{equation}
\begin{equation}
\varepsilon_{\nu}=\frac{1}{K}\sum_{k=0}^{K-1}\frac{\left|\nu_{k}^{(r)}-\nu_{k}\right|}{\nu_{k}}\times100\%.\end{equation}
The recovery errors as a function of the $SNR$ are represented in
Figure 3. The computation was performed by averaging over $M=10^{2}$
cases for each $SNR$ value, using the following parameters: $N=128$,
$K=8$, $T=10^{2}$, $S=K^{2}$. The numerical results show that the
method is robust to noise perturbations, and recovers the spectrum
well up to $SNR\simeq10$, when begins to deteriorate. Also, it is
interesting to note that the error of the complex amplitudes $\varepsilon_{Z}$
grow much faster than the frequency error $\varepsilon_{\nu}$. This
is due to the fact that by increasing the noise perturbation, the
information about the phase of the amplitude is gradually lost, while
the frequencies are still well recovered even for up to $SNR=2$.

\section{Application to Faraday Rotation Measure Synthesis}

Faraday rotation is a physical phenomenon where the position angle
of linearly polarized radiation propagating through a magneto-ionic
medium is rotated as a function of frequency. Recently, the Faraday
rotation measure (RM) synthesis has been re-introduced as an important
method for analyzing multichannel polarized astrophysical radio data,
where multiple emitting regions are present along the single line
of sight of the observations. In practice, the method requires the
recovery of the Faraday depth function from measurements restricted
to limited wavelength ranges, which is an ill-conditioned deconvolution
problem, raising important computational difficulties (see \cite{key-4,key-5} for
a detailed description). 

Faraday rotation is characterized by the Faraday depth (in $\mathrm{rad}\,\mathrm{m}^{-2}$),
which is defined as:\begin{equation}
\phi(r)=0.81\int_{source}^{observer}n_{e}B\cdot dr,\end{equation}
 where $n_{e}$ is the electron density (in $cm^{-3}$) , $B$ is
the magnetic field (in $\mu G$), and $dr$ is the infinitesimal path
length (in parsecs). We also define the complex polarization as:\begin{equation}
P(\lambda^{2})=Q(\lambda^{2})+iU(\lambda^{2}),\end{equation}
 where $I$, $Q$, $U$ are the observed Stokes parameters. The observed
polarization $P(\lambda^{2})$ originates from the emission at all
possible values of the Faraday depth $\phi$, such that:

\begin{equation}
P(\lambda^{2})=\int_{-\infty}^{+\infty}F(\phi)e^{2i\phi\lambda^{2}}d\phi,\end{equation}
 where $F(\phi)$ is the complex Faraday depth function (the intrinsic
polarized flux, as a function of the Faraday depth). Thus, in principle
$F(\phi)$ is the inverse Fourier transform of the observed quantity
$P(\lambda^{2})$:\begin{equation}
F(\phi)=\int_{-\infty}^{+\infty}P(\lambda^{2})e^{-2i\phi\lambda^{2}}d\lambda^{2}.\end{equation}
In general, the number of observations $N$ is limited by the number
of independent measurement channels, and therefore there are many
different potential Faraday depth functions consistent with the measurements.
The usual approach to resolving such ambiguities, is to impose some
extra constraints on the Faraday depth function. Here, we consider
that the complex Faraday depth function is approximated by a small
number of components. More specifically, we assume that $F(\phi)$
contains $K$ (unknown) Dirac components $f_{k}\delta(\phi-\phi_{k})$,
with complex amplitudes $f_{k}\in\mathbb{C}$, and depths $\phi_{k}\in\mathbb{R}$,
$k=0,...,K-1$:\begin{equation}
F(\phi)=\sum_{k=0}^{K-1}f_{k}\delta(\phi-\phi_{k}),\end{equation}
Taking the Fourier transform of $F(\phi)$ we obtain: \begin{equation}
P(\lambda^{2})=\int_{-\infty}^{+\infty}\sum_{k=0}^{K-1}f_{k}\delta(\phi-\phi_{k})e^{2i\phi\lambda^{2}}d\phi=\sum_{k=0}^{K-1}f_{k}e^{2i\phi_{k}\lambda^{2}}.\end{equation}
The goal of the RM synthesis is to find $F(\phi)$ from the observed
values $P(\lambda_{n}^{2})=P_{n}$ (i.e. $Q_{n}$ and $U_{n}$) over
$N$ discrete channels $\lambda_{n}^{2}\in[\lambda_{min}^{2},\lambda_{max}^{2}]$,
$n=0,1,...,N-1$. From the numerical point of view one can apply the
EFSR method directly, by substituting the Faraday depth and the squared
wavelength with: $\phi\leftrightarrow\pi\nu$ and $\lambda^{2}\leftrightarrow t$. 

In Figure 4 we give such an example simulated for the Westerbork Synthesis
Radio Telescope (WSRT) experiment layout \cite{key-4,key-5}. The various parameters associated
with the WSRT experiment layout are listed bellow:
\medskip{}

Wave length range: $\lambda_{min}^{2}=0.639\,\mathrm{m}^{2}$, $\lambda_{max}^{2}=0.905\,\mathrm{m}^{2}$;

The width of an observing channel: $\delta\lambda^{2}=0.0021\,\mathrm{m}^{2}$;

Maximum observable Faraday depth: $\phi_{max}\simeq 800\,\mathrm{rad}\,\mathrm{m}^{-2}$;

Depth space resolution (equivalent to half Nyquist frequency): $\delta\phi\simeq 13\,\mathrm{rad}\,\mathrm{m}^{-2}$;

\medskip{}

We randomly generated $K=5$ sources with complex amplitudes, Faraday
depth $-\phi_{max}<\phi<\phi_{max}$. 
All the components are separated at the Nyquist limit. We also added noise to the generated polarization vector
$P(\lambda^{2})$, such that $SNR=10$. One can see that while DFT fails to recover the 
depth spectrum, while the EFSR method correctly determines the all the components embedded in the
polarization signal, with a small error in the phase, due to the ambiguity
induced by the noise in the data.

\section{Conclusion}

We have presented an exact Fourier spectrum recovery method, which
can be applied in all cases, including those in which the signal contains
frequencies that are not corresponding to the sampling grid. Also,
we have provided an implementation algorithm and we have shown numerically
that the method is also extremely robust to noise perturbations.

\newpage{}
\begin{figure}
\centering \includegraphics{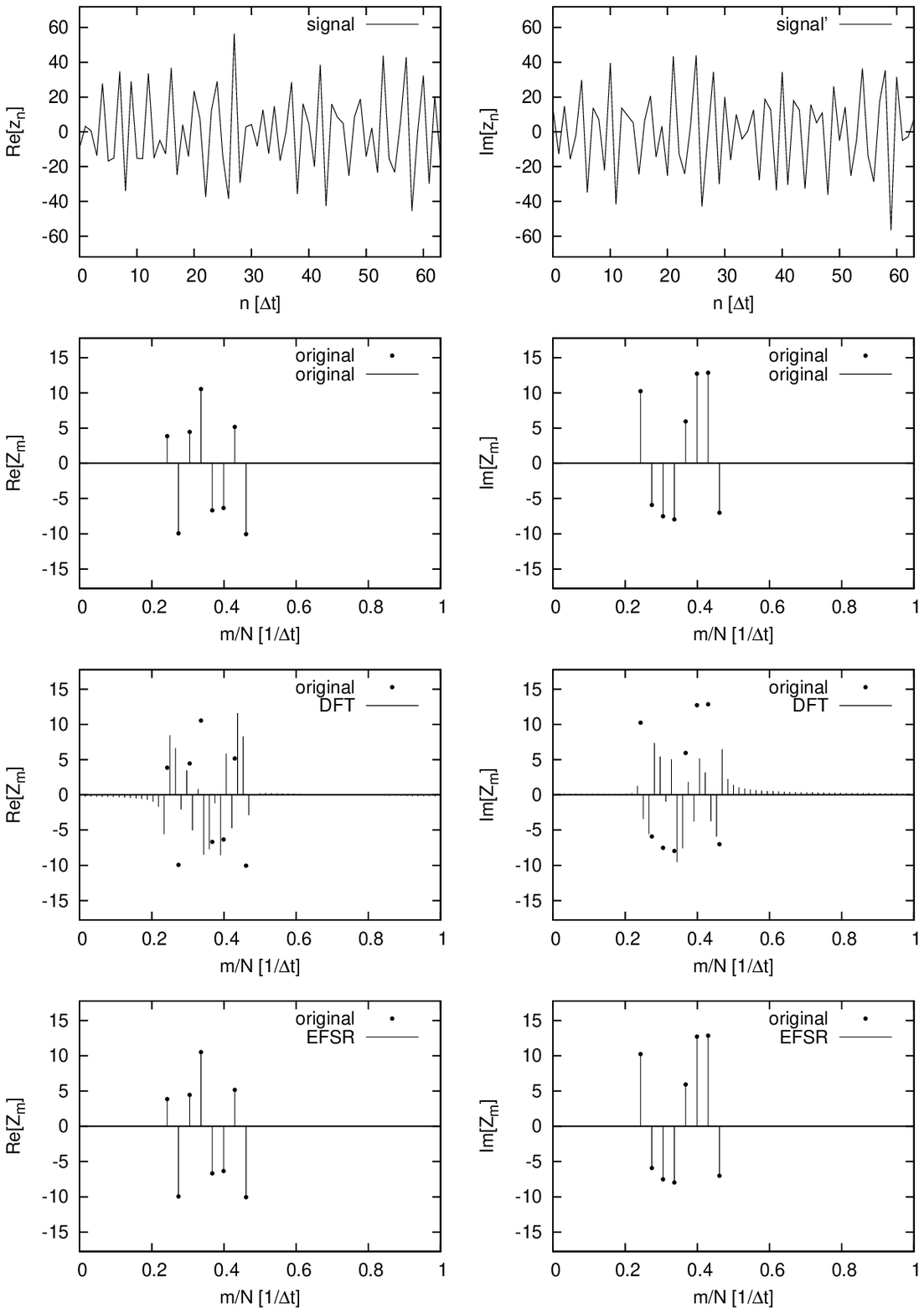}\caption{An example of noiseless signal ($SNR=\infty$), sampled at $N=64$
time steps (1st line), containing $K=8$ Fourier components with frequencies
not falling on the sampling grid (2nd line). The DFT method is failing in this case (3rd line), while the EFSR method recovers
the Fourier spectrum exactly (4th line).}
\end{figure}

\newpage{}
\begin{figure}
\centering \includegraphics{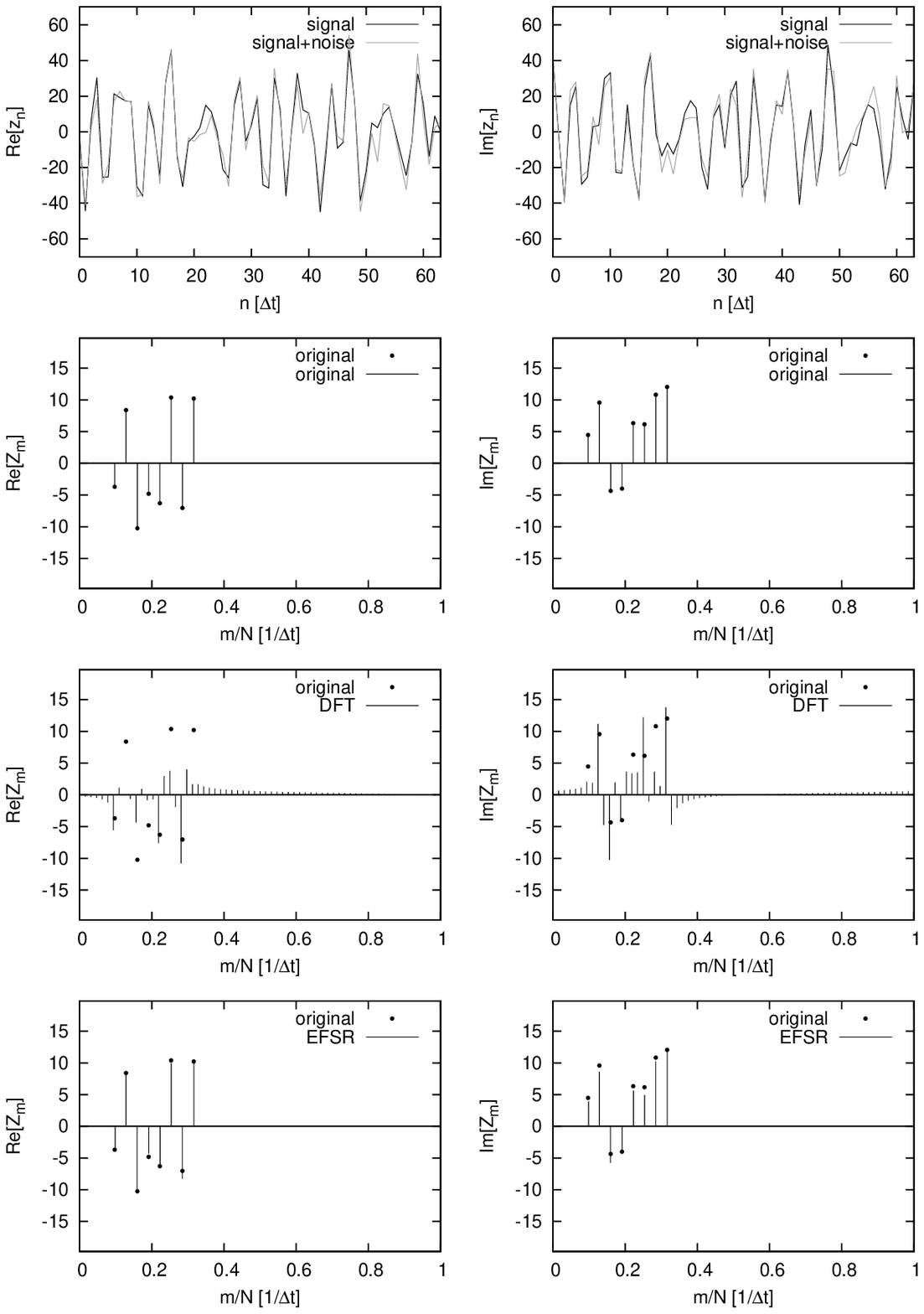}
\caption{An example of noisy signal ($SNR=10$), sampled at $N=64$ time steps
(1st line), containing $K=8$ Fourier components with frequencies
not falling on the sampling grid (2nd line). The DFT method is failing in this case (3rd line), while the EFSR method recovers
the Fourier spectrum almost exactly (4th line).}
\end{figure}

\newpage{}
\begin{figure}
\centering \includegraphics{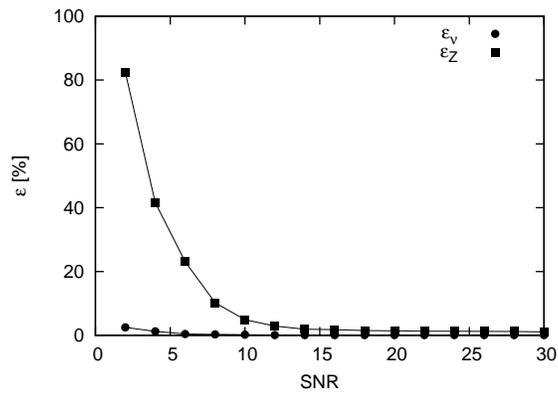}
\caption{Recovery errors as a function of the $SNR$: the error of the complex
amplitudes $\varepsilon_{Z}$ grow much faster than the frequency
error $\varepsilon_{\nu}$.}
\end{figure}

\newpage{}
\begin{figure}
\centering \includegraphics{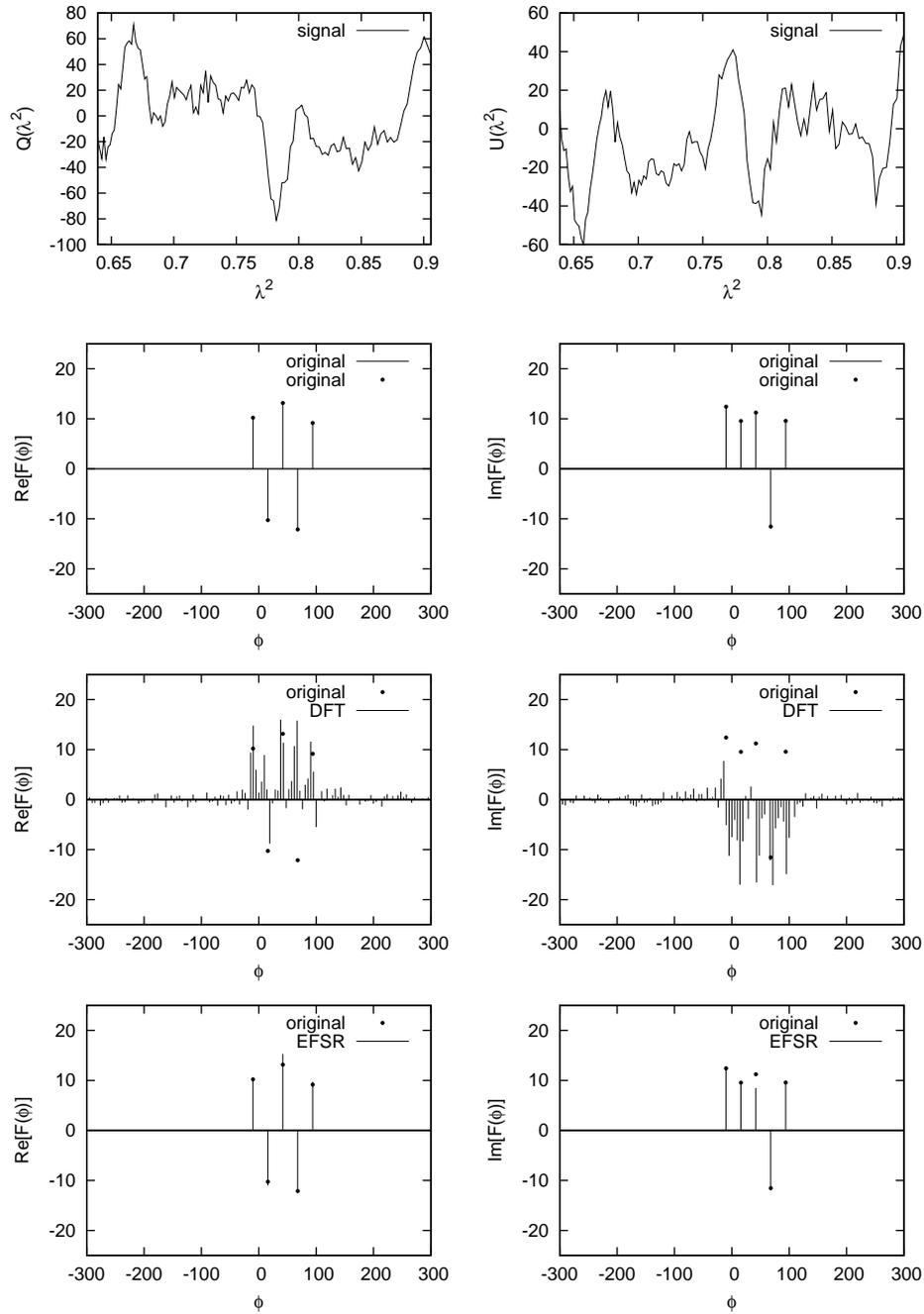}
\caption{An example of noisy rotation measure synthesis ($SNR=10$), for the WSRT experiment layout 
(1st line), containing $K=5$ components (2nd line). The DFT method is failing in this case (3rd line), while the EFSR method recovers
the spectrum almost exactly (4th line).}
\end{figure}

\end{document}